\documentclass[10pt]{IEEEtran}\usepackage[]{graphicx}\usepackage[usenames,dvipsnames]{color}
\makeatletter
\def\maxwidth{ %
  \ifdim\Gin@nat@width>\linewidth
    \linewidth
  \else
    \Gin@nat@width
  \fi
}
\makeatother

\definecolor{fgcolor}{rgb}{0.345, 0.345, 0.345}

\usepackage{framed}
\makeatletter
 {\par\unskip\endMakeFramed%
 \at@end@of@kframe}
\makeatother

\definecolor{shadecolor}{rgb}{.97, .97, .97}
\definecolor{messagecolor}{rgb}{0, 0, 0}
\definecolor{warningcolor}{rgb}{1, 0, 1}
\definecolor{errorcolor}{rgb}{1, 0, 0}

\usepackage{alltt}

\usepackage[utf8]{inputenc}
\usepackage{wrapfig}
\usepackage{tabularx}
\usepackage[T1]{fontenc}
\usepackage{textcomp}
\usepackage[garamond]{mathdesign}

\usepackage{graphicx}
\usepackage{enumerate}

\usepackage[usenames,dvipsnames]{color}
\usepackage[breaklinks]{hyperref}

\hypersetup{colorlinks=true, linkcolor=Black, citecolor=Black, filecolor=Blue,
    urlcolor=Blue, unicode=true}

\usepackage[american]{babel}
\usepackage[numbers]{natbib}

\usepackage{etoolbox}
\makeatletter
\pretocmd{\NAT@citexnum}{\@ifnum{\NAT@ctype>\z@}{\let\NAT@hyper@\relax}{}}{}{}
\makeatother

\def\citepos#1{\citeauthor{#1}'s \citep{#1}}

\usepackage[absolute]{textpos}

\newcolumntype{T}{>{\centering\arraybackslash}m{1cm}}
\newcolumntype{G}{>{\centering\arraybackslash}m{2cm}}
\newcolumntype{V}{>{\centering\arraybackslash}m{3cm}}
\newcolumntype{Z}{>{\arraybackslash}m{5cm}}
\newcolumntype{A}{>{\arraybackslash}m{7cm}}
\newcolumntype{t}{>{\arraybackslash}m{1cm}}
\newcolumntype{g}{>{\arraybackslash}m{2cm}}
\newcolumntype{v}{>{\arraybackslash}m{3.25cm}}
\newcolumntype{z}{>{\centering\arraybackslash}m{4cm}}

\IfFileExists{upquote.sty}{\usepackage{upquote}}{}
\begin{document}

\makeatletter
\def\ps@IEEEtitlepagestyle{%
  \def\@oddfoot{\mycopyrightnotice}%
  \def\@evenfoot{}%
}
\def\mycopyrightnotice{
  {\footnotesize This work has been submitted to the IEEE for possible publication. Copyright may be transferred without notice, after which this version may no longer be accessible. \hfill}
  \gdef\mycopyrightnotice{}
}




\title{Qualities of Quality: A Tertiary Review of Software Quality Measurement Research}

\author{
\IEEEauthorblockN{Kaylea Champion, Sejal Khatri, Benjamin Mako Hill}\\
\IEEEauthorblockA{University of Washington\\
                  Seattle, Washington, USA\\
Email: \{kaylea,sejalk,makohill\}@uw.edu}
}
\date{July, 2021}


\maketitle

\IEEEpeerreviewmaketitle
\begin{abstract}
This paper presents a tertiary review of software quality measurement research. To conduct this review, we examined an initial dataset of 7,811 articles and found 75 relevant and high-quality secondary analyses of software quality research. Synthesizing this body of work, we offer an overview of perspectives, measurement approaches, and trends. 
We identify five distinct perspectives that conceptualize quality as heuristic, as maintainability, as a holistic concept, as structural features of software, and as dependability. We  also identify three key challenges.
First, we find widespread evidence of validity questions with common measures. Second, we observe the application of machine learning methods without adequate evaluation. Third, we observe the use of aging datasets.
Finally, from these observations, we sketch a path toward a theoretical framework that will allow software engineering researchers to systematically confront these weaknesses while remaining grounded in the experiences of developers and the real world in which code is ultimately deployed.
\end{abstract}

\section{Introduction}
\label{sec:introduction}
In that our day-to-day lives depend on an ecosystem of software components serving as global digital infrastructure, software quality is of immense concern to both software engineering and associated disciplines, and to society in general \citep{eghbal_roads_2016}.
In order to understand the risks posed by low quality software infrastructure and to inform attempts to improve software quality, we embarked on this project to answer two fundamental questions about software quality: 

\begin{itemize}[
    \setlength{\itemindent}{\dimexpr\labelwidth+\labelsep}
    ]
    
    \item[{\textbf{RQ1:}}] How is software quality defined---what do we mean when we talk about high or low quality software? 
    
    \item[{\textbf{RQ2:}}] How is software quality measured?
\end{itemize}

Given the existence of a large number of secondary reviews on software quality measurement, we conducted a tertiary review---that is, a review of reviews. 
We begin by reviewing the insights of past tertiary review efforts in §\ref{sec:background}, before offering a description of our tertiary review method and dataset in §\ref{sec:method}. We offer
a synthesis of the secondary reviews we identify in the form of five perspectives on quality in §\ref{sec:synthesis}. 
We report the results of our analysis in a quantitative results section in §\ref{sec:results} and comment on these findings in §\ref{sec:discussion}. We detail potential threats in §\ref{sec:threats} before concluding in §\ref{sec:conclusion} with implications for software engineering researcher.

\section{Related Work}
\label{sec:background}
In 2004, a highly-cited piece from \citeauthor{kitchenham_evidence-based_2004} called for software engineering research to take up an evidence-based approach, inspired by evidence-based medicine \citep{kitchenham_evidence-based_2004}.
\citeauthor{kitchenham_evidence-based_2004} describe evidence-based software engineering as tackling not only carefully-defined research questions and developing strong evidence, but also ``critically appraising that evidence for its validity ... impact ... and applicability'' (p. 275), integrating this critical appraisal into findings and future work, and continuing to evaluate past work for improvement. Our review is a response to this challenge. 

A series of previous reviews have attempted to sketch the contours of a still-emerging body of research on software quality in various ways. Previous reviews have tended to focus more narrowly on specific challenges with respect to evidence and validity as well specific promising approaches. Our review is complementary to these prior reviews in that it grapples with software quality in a cross-perspective manner and in that it is broader both in scope and scale. 

With respect to evidence, the mapping review in \citet{bailey_evidence_2007} found an absence of comparative evaluation of the quality of software produced through Object-Oriented versus non-OO paradigms. With respect to validity, \citet{syeed_prediction_2014} found in their systematic review of contributions from software quality research to open source development practices that many of the most significant quality predictors were rarely used in research. 
Prior authors have also identified potential avenues for progress: mining software repositories unlocks longitudinal data for quality research \citep{de_farias_systematic_2016}, and AI/data science techniques have much to contribute if data and metric validity concerns can be addressed \citep{fernandez_del_carpio_techniques_2018}. 
We found one previous tertiary analysis in the field of software quality: \citet{elmidaoui_software_2017} examined nine secondary studies of software maintainability as a specific facet of quality and found that while maintainability prediction is an active area of work, model performance and validation continue to be a concern.

Our paper offers three contributions to the software engineering literature on quality through the examination and synthesis of 75 secondary reviews.
First, we develop an ontology of five perspectives used to define and measure software quality in current lines of research. 
Second, we describe trends and topical coverage which stretch across the software quality research field. 
Third and finally, we offer an analysis of the key victories and most vexing struggles currently facing software quality researchers.

\section{Review Methodology}
\label{sec:method}

\begin{figure}[t]
\centering
\includegraphics[width=0.35\textwidth]{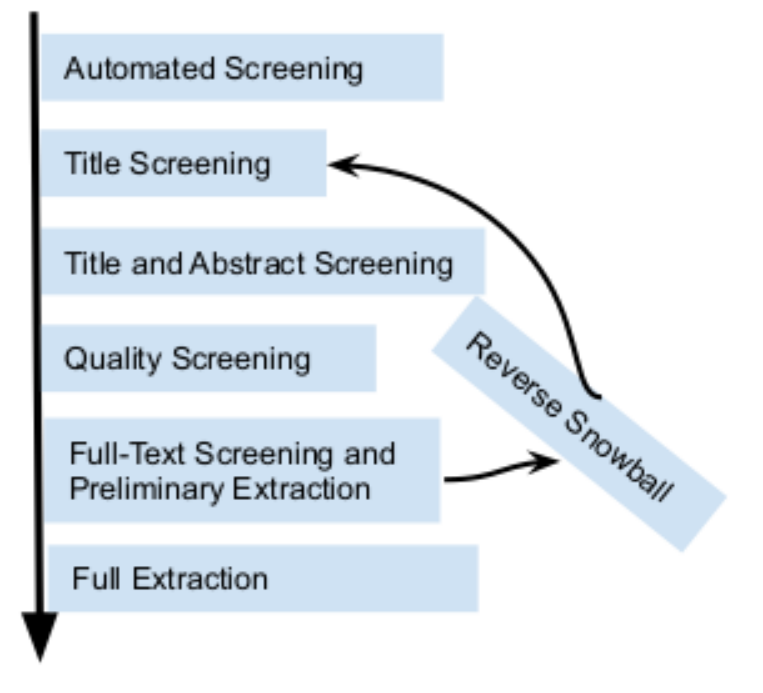}
\caption{Stages of the systematic literature search and screening process we conduct}
\label{fig:process}
\end{figure}

As a systematic literature review, this tertiary analysis follows a reproducible process for identifying, analyzing, and synthesizing a comprehensive set of relevant high-quality research. Just as secondary reviews treat primary studies as their evidence, tertiary reviews use secondary reviews as inputs. 
Below, we describe our use of the systematic review method described in \citet{kitchenham_evidence-based_2004} using the stages illustrated in Figure \ref{fig:process}. When we refer to papers that are part of the corpus analyzed through systematic review, we use a separate index number, prefixed by `S' (i.e., S1-S75). These papers are all listed in this review in the Secondary Literature section. Other citations are included as normal.

Our review protocol was informed by guidelines articulated for software engineering, information systems, and medicine \citep{kitchenham_evidence-based_2004, smith_methodology_2011, okoli_guide_2015}. While conducting this review, the review team met regularly to verify fidelity to the protocol and to update the search plan when needed. We adopted shared processes around search and retrieval including the use of shared spreadsheets for data capture and the use of a shared reference database using Zotero.

\subsection{Automated Screening}

\begin{table}
\caption{List of numbers of collected papers for all indices and sub-indices}
\begin{center}
\begin{tabular}{p{2cm} p{3.5cm} p{2cm}}
Index & Sub-Index & Results (raw) \\
\hline
EBSCO & Academic Search Complete & 870\\
Proquest & Sociology Abstracts & 198 \\
Web of Science & All Databases &  2216 \\
Scopus &  & 5867 \\
Inspec/Compendex & & 765  \\
\hline
Total & & 9916  \\
\end{tabular}
\end{center}
\label{tab:indexResults}
\end{table}

\begin{table}
\caption{Search keywords}
\begin{center}
\begin{tabular}{l l v}
Term & Field & Synonym\\
\hline
systematic review & abstract & literature review, systematic mapping, meta-review \\
quality & abstract & bug, flaw, vulnerab*, problem, reliab*, stab*,  performance, issue \\ 
software & subject & --- \\ 
\end{tabular}
\end{center}
\label{tab:terms}
\end{table}

We collected our initial dataset from a set of database indexes of scholarly papers using a schema of keywords, operators, and fields. 
The search process was developed in collaboration with research librarians with expertise in Computer Science and in Communication. We ensured that publications by the ACM and IEEE were fully indexed by these services. The final list of databases indices used and counts of results from our search process is listed in Table \ref{tab:indexResults}. Because software quality is a subject of active debate and consideration, we refined the list of keywords iteratively, ultimately using those in Table \ref{tab:terms}.
Indices vary in their search capabilities; we include the specific queries we ran in each database as supplemental material in our dataset and code release.\footnote{\url{https://doi.org/10.7910/DVN/XPOUO4}} 
To manage the scope of the project, we limited our search to work published in the years 2000 to 2020 inclusive and written in English. 
Our initial search results returned  9,916 results.

As S22 and S56 found, index-only search techniques are often insufficient to find all literature related to a question. As a result, we supplemented our primary results using the reverse snowballing technique \citep{wohlin_guidelines_2014}---i.e., when the secondary works we identified referenced other secondary studies, 
we screened those as well. 
We repeated the snowball process three times, stopping when we found no references to any new secondary work. 
The reviews found through snowballing were filtered in the same manner as the results from our index searches. 

We conducted both an automated and a manual inspection process to identify and remove all duplicates. This resulted in a total of 7,811 articles in our initial deduplicated dataset.

\subsection{Title, Abstract, Quality, and Full-Text Screening}

\begin{table}
\caption{Screening Criteria}
\begin{center}
\begin{tabular}{l A}
Criteria & Description \\
\hline
in scope & must be a secondary analysis; must define software quality; must attend to measurement of quality\\
generalizable & must offer insight for a range of quality measurement tasks; exclude articles that are narrowly applicable to a single domain \\
high quality & must be peer reviewed; not gray literature, self-published, or publications from predatory journals \\
\end{tabular}
\end{center}
\label{tab:qualityusageaq}
\end{table}

\begin{table}
\caption{
Articles present after screening stages}
\begin{center}
\begin{tabular}{l r r r r r r} 
Round & Initial & Unique & Automated & Title & Abstract & Final \\
\hline
Primary 
& 9916 & 7656 & - & 627 & 194 & 54 \\
Snowball 1
& 227 & 106 & 93 & 88 & 80 & 18 \\
Snowball 2
& 87 & 48 & 37 & 36 & 33 & 3 \\
Snowball 3
& 4 & 1 & 1 & 1 & 1 & 0 \\
Total  
& 10234
& 7811
& 131
& 752
& 308
& 75
\\
\end{tabular}
\end{center}
\label{tab:screenres}
\end{table}

After deduplication, we screened reviews based on their content and quality using the three inclusion criteria listed in Table \ref{tab:qualityusageaq}.
Screening was conducted in four rounds, reviewing each papers' (1) title, (2) title and abstract, (3) quality, and (4) full text.
At each step, we made conservative decisions and retained a paper for the next round of deeper review if we were uncertain. The results of each screening step are summarized in Table \ref{tab:screenres}.

Our inclusion criteria for quality were modeled after \citepos{kitchenham_systematic_2009} and intended to limit our sample to peer-reviewed publications. 
We removed 11 items published through predatory publishers which we determined by examining whether a given journal appeared on Beall's list.\footnote{\url{https://beallslist.net/}} We did so because predatory publishers do not engage in substantive peer review \citep{djuric_penetrating_2015}. We also excluded one paper which appeared only in a preprint form
and five items that were PhD theses. 

We examined conference papers with care and searched for associated journal articles in each case because it is common for preliminary results to be presented at a conference, revised, and then later published in a journal. In the 13 occasions when a journal article was found with the same authors and general findings as a previously published conference paper, we excluded the conference paper and retained the journal article to prevent redundancy.

The process resulted in a dataset of 75 secondary reviews of software quality. We summarize the results of the search, screening, and the snowballing process in Table \ref{tab:screenres}. 
Bibliographic data for the full list of 75 reviews used in this paper is available in the secondary bibliography. The full dataset is available in the supplemental materials.

\subsection{
Synthesis Through Thematic Analysis}

We conducted a thematic analysis of the 75 reviews in our final corpus following the qualitative research method described by \citet{braun_using_2006}. Thematic analysis creates a bridge between qualitative and quantitative results through the analysis of the overall argument, approach, and assumptions made in a work, as well as the specific and countable traits of a review. Our thematic analysis was focused on understanding how each review defined and measured software quality and how these decisions shaped the authors' results and takeaways. The output of our thematic analysis is a series of categories that capture the approaches being taken across the field.
Preliminary data extraction was done to a spreadsheet, shared with the team, which assigned a unique ID to each study. We recorded the source (e.g., original search, snowball round), publication venue, abstract for quick reference, the date of publication, and the author's name for their overarching method (systematic mapping, systematic literature review, etc.). 

To perform our thematic analysis, the first author read each review in detail and summarized the methods, data, findings, and concerns expressed in each one. The first author weighed multiple categorization schemes and common threads across the work, re-reading each article as necessary. This process culminated in an iterative series of proposals made to the other two authors for ways to describe how the perspectives expressed in the reviews might be grouped and distinguished from one another. Reviews were sorted and re-sorted until we arrived at five themes that characterize differing perspectives on the subject of software quality. As part of the synthesis process that followed, we also identified opportunities and concerns that recur across varying definitions of quality and seem to characterize the state of software quality research as a subfield; these observations are included in §\ref{sec:discussion}. 

\section{Findings: Five Perspectives on Quality}
\label{sec:synthesis}

We identified five perspectives on quality through our thematic analysis:
\textit{quality is heuristic} for reviews applying and seeking to automate the detection of practice-oriented warning signals, antipatterns, and ``bad smells'' in code (§\ref{sec:d_clean});
\textit{quality is maintainability} for reviews examining time-based processes as they apply to code, such as decay and maintenance (§\ref{sec:d_time});
\textit{quality is holistic} for studies taking up a broad, common-sense, or multi-faceted perspective (§\ref{sec:d_holistic});
\textit{quality is structural} for reviews approaching quality as a matter of predicting areas prone to bugs and defects (§\ref{sec:d_bugs}); and 
\textit{quality is dependability} for reviews approaching quality as the reliability and robustness of code (§\ref{sec:d_depend}); 

\begin{table}
\caption{Reviews in each theme}
\begin{center}
\begin{tabular}{l r r}
Theme & Count & Proportion \\    \hline Quality is Heuristic &  20 & 27\% \\    Quality is Maintainability &  19 & 25\% \\    Quality is Holistic &  18 & 24\% \\    Quality is Structural &  10 & 13\% \\    Quality is Dependability &   8 & 11\% \\    
\label{tab:byTheme}
\end{tabular}
\end{center}
\end{table}

The distribution of reviews across these five perspectives is shown in Table \ref{tab:byTheme}. 
Although our categorization scheme is based on our own analysis of the reviews, several categories we identified are similar to those used by S33 \citep{ruiz_measuring_2011}. There is also some overlap between the themes we identify and the ISO standards for software quality that express two key facets: quality in requirements conformance and quality in use \cite{bevan_quality_1997}.
We present our synthesis of the work done within each perspective in turn by describing the theme, summarizing key findings, and then detailing challenges and future areas of work.

\subsection{Quality Is Heuristic}
\label{sec:d_clean}


This theme gathers work taking the perspective that readable, well-designed code is easier to maintain and less likely to cause future issues. The primary strategy for improving quality within this theme is refactoring: revising code in order to improve the underlying design without altering overall functionality. All of the 20 review articles we associated with the theme of heuristics for clean code either directly referenced Fowler's 1999 book \textit{Refactoring} \citep{fowler_refactoring_1999} and/or discussed the notion of code smells. Fowler describes Kent Beck's turn of phrase ``code smells'' as ``not-quite-right code:'' a series of tell-tale signs of code problems, each with a conceptual and evocative name (e.g., God Class). Fowler characterized these smells as inherently subjective but relatively easy for inexperienced programmers to identify. Fowler also  associated each bad smell with a proposed set of refactoring strategies that could remedy the issue. 

Although some of the empirical work reported by these reviews also propose new smells, research on this topic typically addresses Fowler's list of smells (especially two of Fowler's original 22: Duplicated Code and Large Class) [S41]. The earliest work in this theme used manual identification techniques, but machine learning (ML) techniques are increasingly prevalent. This includes lines of work seeking to identify optimal algorithms and cutpoints and others seeking automation of Fowler's associated refactoring strategies. 

Substantial effort has gone into the refinement of code smell detection techniques. S30 provides a review of code smell mining approaches including static, dynamic, semantic, metrics, and historical analysis, using structural features in service of a quality as heuristic perspective. S30 also enumerates several detection techniques in use, with disadvantages noted in parentheses: manual (time-consuming and error-prone); metric-based (no consensus on thresholds); symptom-based (low precision due to varying interpretations of symptoms); probabilistic (no disadvantages listed); visualization-based (about 50\% precision and recall), search-based techniques using ML (success varies); and cooperative-based using genetic programming (may not generalize). 

Our analysis found substantial disagreement in the literature on whether bad code smells are a problem for quality.
We found that some reviews assumed that code smells were negatively associated with quality and focused on assessing empirical work oriented to detecting smells [S12, S14, S75]. Other reviews examined the weight of evidence associating bad smells  with low quality software [S4, S36, S41, S42]. This latter group found that ``bad'' smells had only a weak, and sometimes even positive, relationship with measures of software quality.  
These results suggest that future researchers should not assume that the presence of bad code smells is an indicator of lower quality. Instead, researchers should evaluate the evidence relating each specific code smell heuristic they are using for their measurement to a specific facet of quality. 

Recognizing that code smells are both widely used and unreliable indicators of quality, S59 took up the task of categorizing false positives in antipattern and code smell research. S59 characterized the evidentiary challenge as two-fold: a practical challenge due to the fact that manual validation of results from code smell analysis can be time-prohibitive, and a construct validity challenge caused both by a limited understanding of what maintainable code looks like and a lack of empirical support. 
S5 found that ML based studies using metrics associated with smells performed better than ML studies that used smells directly. However, this result was uncertain given weak evaluation metric reporting---S5 observed that studies more often reported threshold-dependent metrics 
instead of metrics such as AUC. 

S42 expressed concern about the relative predominance of academic researchers in this area. While S34 found that all authors in the 78 studies they examined were from academic environments, they were ``mainly working in research groups with support from industry.'' This more nuanced view of affiliation led S34 to conclude that there is no clear answer as to whether research on smells is primarily conducted by industry or academia.

Many researchers called for better coordination between research and practice. 
In addition to addressing basic validation challenges, future work taking the quality as heuristic perspective may find substantial value in learning from practitioners directly. S52 found that many refactoring strategies commonly used by practitioners have not been investigated 
while some rarely-practiced approaches have been the object of substantial study. 
Because code smells are a reflection of practitioner experience, maintaining close awareness of practitioner experience seems likely to support more applicable results. 


\subsection{Quality is Maintainability}
\label{sec:d_time}



The perspectives in the 19 studies we discuss in this section describe quality in relation to the process of software maintenance. We identified four distinct lines of work within this broad theme: (1) studies exploring the broad notion of \textit{maintainability in general}; research on \textit{evolvability} as a measure of quality; research on \textit{technical debt};
and work seeking to understand processes of \textit{aging} in software.

\subsubsection{Maintainability in general}
Maintainability as a facet of software quality is concerned with how efficiently and effectively a given codebase can be maintained over time. Despite the fact that multiple studies have found that maintenance is the most costly portion of a software product's lifecycle, S31 and S72 observed that very little is known about predicting maintainability. 
S73 observed a lack of consensus with respect to open source maintainability metrics but that coupling metrics were most common, appearing in 5 of the 14 studies they identified. 

S31 recommended more robust scales for quantifying maintainability and emphasized the need for model validation. In particular, S31 pointed out that only 4 of the 15 studies they identified used a validation technique like leave-one-out (LOO) cross validation. S54 examined maintainability measurement studies in the range of 2003-2012 and also found that external validation was lacking. 

S72 suggested that developer team performance-based metrics like Mean Time to Repair, Mean Corrective Maintenance Time, and so on, were appropriate for measuring maintainability. 
S72 found that few studies had considered the maintainability of non-OO systems, the impact of Agile processes, or component-based development.

\subsubsection{Evolvability}
S29 framed the evolvability challenge as a kind of ``stability''---that is, whether a code base can be adapted to incorporate new needs with minimal effort rather than needing to be disruptively or deeply overhauled. 
Measurement of evolvability as a facet of quality is in a formative phase. As a result, reviews categorized with this theme used a variety of terms including evolvability, stability, and reusability. 
S11 examined architecture quality studies as they relate to evolvability, 
but found most were case studies. Of the 82 studies they examined, only 10 were metric-oriented studies with no common or externally validated measures among them. 
S29 observed that while multiple evolvability metrics have been proposed and validation work is needed, the metrics share an emphasis on the fundamental design principles of low coupling, high cohesion, and functional independence.
S35 conducted an exhaustive review of the ``stability'' concept at the design, architecture, code, and requirements level across 166 articles.
Although the quantity of work done in this area is increasing, S35 found that stability studies have not generally been conducted as empirical work in applied settings to understand stability successes and failures in practice. .

A related concept is ``reusability'' which is a sub-facet of maintainability in the ISO25010 quality standard. S10 considered quality as reusability and sought to determine which of a long list of non-functional requirements (NFRs) were loosely related to, positively correlated with, or negatively correlated with, reusability.
The tradeoffs among quality facets observed in the articles reviewed in §\ref{sec:d_clean} were also evident in S10's review of quality as reusability: of the quality facets outlined in ISO25010, reusability was negatively correlated with 4 of the 47 considered (performance, dependability, cost, and, surprisingly, portability) and positively correlated with 16 of the 47 facets (including safety, interoperability, and documentation). 

\subsubsection{Technical debt}

Technical debt is a metaphor used in software engineering since at least 1993 when \citeauthor{cunningham_wycash_1993} wrote that ``shipping first time code is like going into debt. A little debt speeds development so long as it is paid back promptly with a rewrite'' \citep[p. 30]{cunningham_wycash_1993}.
Given the descriptive success of the technical debt metaphor, S68's 2012 article set out to establish a theoretical model using a systematic review. S68 found that technical debt encompasses a range of concepts and is involved with examples including lack of abstraction and code duplication, excessive complexity, and deviation from a reference architecture. S68 also found that the definition of technical debt varied across the software engineering literature and was used to invoke known bugs, incomplete testing, missing features, scant documentation, and more. S68 proposed a theoretical framework that elaborates the metaphor into such concepts as interest, bankruptcy, and gearing (also known as ``leverage'' or sustainable debt level). S68's elaboration of the concept includes such elements as code decay, design/architectural debt, infrastructural debt, documentation debt, testing debt, known issues, and missing features. 
S68 identified potential causes of technical debt as project constraints, lack of debt visibility, reckless decisions, and deliberate (or inadvertent) debt accrual. Potential consequences of these forms of debt include improvement or risk to timelines, maintenance difficulty, and ongoing quality problems. 

S58 extended S68's analysis to incorporate practitioner perspectives, expanding the metaphor further to incorporate the very real cost of technical debt and to include the notions of different kinds of decision-making processes associated with debt (strategic debt, tactical debt, and incremental debt), debt amnesty and debt repayment, as well as the morale and productivity impacts.
S53 offered several additional categories of debt such as people debt, process debt, build debt, service debt, usability debt, and versioning debt.

S22 observed that although the term technical debt was coined in the early 1990s, the recent growth in published studies of technical debt did not begin until 2009. S22 speculated that the Managing Technical Debt workshop may have raised awareness of this research avenue. S22 found that empirical studies of technical debt differ in their definitions of debt, especially with regard to whether defects should be considered technical debt. Although the concept of technical debt has been broadly elaborated, S22 found that most technical debt research was focused specifically on code technical debt, especially the measurement and identification of it, although architectural and design debt have also received some attention. 

Two secondary studies we identified specifically addressed architectural debt. S8 enumerated types of architectural technical debt, described refactoring strategies and proposed a unified model of architectural debt. S8 found that no tools tackled the full spectrum of architectural technical debt and its management. 
S47 observed that the bulk of architectural technical debt studies were conducted from an academic perspective and that more collaboration with industry could lead to beneficial knowledge exchanges.

While most authors articulated technical debt as a concept distinct from code smell, S41 described them as synonymous. 
In any case, the concepts of debt and smell have a complex relationship: a bad-smelling section of code may well represent substantial technical debt. However, some debt may not smell and some smells may not be debt sources but rather, say, an intentional design choice that improves performance. S75 described design smells as a leading source of technical debt. 
Smells and debts may also share an ultimate cause: S14 lists lack of skill or knowledge, changing requirements, technology constraints, process problems, time pressures, and politics as precursors of code smells. These same contingencies, constraints, and limitations are also implicated in technical debt, particularly code-level and design-level debt. 

S22 found that the most widespread approach to measuring technical debt is a series of calculation models, although some approaches also use code-level metrics. S53 found that the most common indicator of technical debt is the code smell, with common measures including detection of the God Class code smell, measures of code complexity, and searches for duplicated code. S22 found that, as with code smells, the most common approach to repayment of technical debt is refactoring, and to a lesser extent full rewriting, re-engineering, or automation. 
S22 identified 29 different tools for tackling technical debt including both products from major vendors and numerous free and open source options. S22 found that these tools typically tackle code and design debt using source code as input, but that none work to prevent technical debt. 
S32 and S56 took up the task of developing decision criteria for the question of repaying technical debt. 
Once again, evidentiary challenges persist:  none of the 38 articles reviewed in S32 and only 4 of the 100 articles reviewed in S53 include any empirical work. 


\subsubsection{Software Aging}

Software aging studies typically use one of two definitions: they either define aging as a consequence of code flaws that come to light in a long-running system (e.g., memory leaks) that need rejuvenation (e.g., through a reboot) or else they define aging as a property of code bases, that emerges as maintainers diverge from original requirements and architecture. It should be noted that these two definitions, although using the same terminology and referring to change over time, are substantively different. Addressing the needs of long-running systems is a relatively narrow perspective with a focused set of concerns, while divergence from an original state is a broad and ongoing consideration. 

Approaching aging from the narrow long-running-systems perspective, S57 finds that most models for studying aging are Markov-based---i.e., they model the likelihood of an aging failure based on recent history and randomization. However, measurement-based approaches taking advantage of time-series analysis on aging indicators from system-level variables are also in use. ML-based techniques using large numbers of system-level variables and threshold-based approaches which monitor aging indicators and trigger a rejuvenation action when the threshold is surpassed are also common. 

S57 emphasized a need to connect aging and rejuvenation studies to real systems, including safety-critical systems. This is especially important for techniques that allow detection of aging problems during software design and validation rather than requiring long run-times. There appears to be some progress with respect to connecting this line of work to practitioner needs. S22 found that 
although aging and rejuvenation work in the 1990s was purely analytical, 
the proportion of studies engaged in empirical or analytical-empirical hybrid work has increased substantially since then.
S46 identified embedded systems as an important target for future research.
This field is also responsive to the existence of dedicated publication venues: S46 observed that publications in this topic were generally increasing but that in a year when the Workshop on Software Aging Research was not held, the number of relevant publications plummeted to 1, from 15 the year prior and 25 the year after.

Taking up the latter definition of aging as divergence from an original state, S6 places aging under the heading of code decay. For S6, code decay is comprised of architectural flaws (including smells, erosion, and general decay), design/code flaws (also including smells, erosion, and general decay), and software aging. 
S6 enumerates the many metrics used for decay detection in the literature, including increases in change span, increases in the number of classes, increases in the values of coupling metrics (e.g., the average number of classes used per class in a package, called coupling between objects, or CBO) between versions, increases in the number of God Classes, Data Classes, or Brain Classes, increases in the number of code smells (e.g., Shotgun Surgery, Feature Envy), as well as increases in the number of design rule violations. 

\subsection{Quality is Holistic}
\label{sec:d_holistic}

We found 18 secondary studies that examined quality from a holistic or multi-faceted perspective. These studies either invoke broad or common-sense notions like system success, or referenced multifaceted structured quality standards like those developed by the ISO.  Two of the reviews we identified are broader syntheses that include a section on quality [S51, S67]. Five of the reviews we identified took up the task of quality definition, prediction, and modeling [S44, S33, S26, S3, S74]. Seven worked to identify specific metrics associated with a broad view of quality [S21, S24, S37, S45, S48, S55, S63]. Finally, four take up the particular question of whether user participation in development influences overall quality [S2, S7, S61, S65].


S33 and S67 sought to compare the relative quality of software based on differences in software production processes. In particular, they asked if the open code base and interest-driven software development process found in peer production/open source would be associated with better overall quality. 
S67 found that openness of code alone is not enough to produce quality: developer commitment (and action) is also required. 
S33 found that empirical work on this topic reached divergent conclusions depending on the measures and evaluation criteria used, suggesting that the answer to the question ``is open source better?'' is largely contingent. 

Another common question animating holistic measurement studies asked: Does involving end users in system development increase the probability of success? Unsurprisingly, the answer based on these reviews is again contingent on measurement approach. If success is measured by users' attitudes toward the resulting software, then the answer seems to be yes. However, if success is measured by productivity outcomes, then the effect is inconsistent and smaller if present at all [S61, S65, S7]. That said, more recent studies tend to focus on attitudinal measures and HCI notions like ease of use, while older studies focus on productivity [S2]. This leaves open the question of whether studies of more recent development projects would now find productivity gains associated with end-user involvement. 

S33 and S3 considered the applicability of holistic models to open source and recommended that measures of quality should incorporate factors of the production community as well factors of the software artifact. 
However, S51 and S67 both found that most research on open source software quality focused only on artifacts. In S63's systematic review of quality metrics, 79\% of the metrics they identified concerned software products, 12\% operated at the software project level, and 9\% concerned software processes. 

The holistic perspective on quality found in secondary literature was not always well-served by the empirical literature on which it was built. S67 found that while many authors proposed holistic and multidimensional measures of quality, most empirical studies only took up a single dimension.
S26 found that relatively few studies made use of multi-dimensional software quality models and advocated greater attention to the ISO 25010 or ``SQuaRE'' standard when considering software quality. 
This suggests substantial empirical work remains to be done, including holistic comparative studies incorporating variables from both artifactual and process-level perspectives.

\subsection{Quality Is Structural}
\label{sec:d_bugs} 

Of our 75 identified secondary studies, 
ten 
examined quality through the lens of predicting defect-prone areas of code. This approach can optimize the development process by alerting developers to issues early, and can assist QA engineers in targeting the most fault prone areas for more intensive scrutiny. 

Many studies in this theme used static metrics extracted from Halstead \citep{halstead_elements_1979}
and McCabe \citep{mccabe_complexity_1976} 
[S15, S25] with OO bug prediction studies tending to use the Chidamber-Kemerer (C\&K) \citep{chidamber_metrics_1994} method-level metrics, although several other sets of metrics have been proposed 
 [S15, S62, S28]. 
S43 reviewed dynamic metrics, which allow for measurement to occur in a contextualized running system. Because dynamic metrics generally extend the essential logic of static metrics such as C\&K, they are at risk of inheriting any validity problems present in the underlying static metrics. However, dynamic metrics also allow for the study of runtime traits such as utilization of polymorphism. 
S28 found that despite this proliferation of metrics, relatively few had been validated at scale (with the exception of C\&K). S23 encouraged extending the use of class-level metrics because of their applicability during the design phase.
Despite this heavy use of code-oriented metrics, S28 found that process-oriented metrics such as code delta, code churn, code age, and change set size were stronger predictors of post-release bugs than static code metrics. 

S20 engaged with the relatively new topic of cross-project defect prediction---i.e., the notion that ML algorithms might be trained on one project and then used to assess another. S20 observes that although many ML algorithms ``work well under the assumption that source and target data are drawn from the same distribution, [this assumption] might not hold for cross-project defect prediction'' (p. 124). 
Several studies summarized by S20 found that prediction may not be bi-directional: models trained on open-source code had high accuracy when applied to closed-source code, but the reverse was not true. 
S27 argues that the ability to predict bugs suggests that early intervention may be possible, but that more work is needed to apply strong empirical methods to identify effective metrics to address early parts of the software lifecycle. S15 and S23 called for research into class-level metrics to enable fault prediction early (i.e., during the design phase). S28 found that static metrics were of limited utility in predicting fault-proneness in highly iterative development processes. 
Another area of concern shared among defect prediction scholars is the connection between academic research and applied and industrial environments. S20 expressed concern that defect prediction research continued to be conducted in laboratory environments rather than in the context of practitioners' work. 
S28 remarked that academic researchers conduct most defection prediction research and that academics were more likely to consider OO metrics while industry researchers tended to use process metrics. Given S28's finding that the process metrics used by industry researchers were also more effective, they recommended that academic researchers follow the lead of industry researchers in this regard. 

\subsection{Quality Is Dependability}
\label{sec:d_depend}
Finally, we identified eight 
reviews that defined quality as dependability and robustness---i.e., the consistent operation of software despite the occurrence of faulty input or irregular operating conditions.
Most robustness studies targeted design/implementation and verification/validation development phases rather than considering the requirements engineering and maintenance processes [S39].  
S70 identified some reliability-specific process and product-level metrics in use: number of defects, 
defect rate, 
mean time to failure, 
mean time to repair, 
and transaction time.
The authors of S39 expressed concern at the lack of empirical validation work in this topic, a finding consistent with S70. Most studies were solution proposals and tended to emphasize method; 65\% had weak or no evaluation. S39 and S17 observed that a majority of papers were academic experiments, conducted in lab settings, with relatively few examining real-world case studies such as open source or industrial deployments. S50 also observed a disconnection from practice and a tendency toward theoretical work, observing that a lack of public experimental datasets may be partially to blame. 

S19 examined 27 studies of testability and robustness and found that testability analysis offers multiple types of insight into robustness. Testability incorporates error propagation analysis, assessment of exception handling, and implementation of strict rules for design and implementation. 
S40 examined studies developing and evaluating metrics for testability as a facet of software quality and found that the most common metrics in use considered traits of observability (the ability to determine if an input influences output) and controllability (the ease of producing a specific output given a specific input). 

Future researchers may find the research taxonomies proposed in S17 and S50 to be useful. While both used keyword aggregation, S17 focused on a taxonomy of modeling approaches while S50 proposes a taxonomy that includes research into both testing, assessment, and analysis as well as modeling. 

Yet again, also we observed concern about the measures in use: S39 expressed concern at the lack of empirical validation work observing that 65\% of the articles they reviewed reported weak or no evaluation. Instead, most studies were solution proposals and tended to emphasize method, a finding consistent with S70. Several authors also commented on the lack of connection between practice and theory. S39 and S17 observed that a majority of papers were academic experiments, conducted in lab settings, with relatively few examining real-world case studies such as open source or industrial deployments. S50 also observed a lack of direct relevance to practice and a tendency toward theoretical work, suggesting that a lack of public experimental datasets may be partially to blame. 
S38 observed a lack of sufficient detail about methods being used for measurement. Only 25\% of the security evaluation methods they reviewed included sufficient detail for the method to be applied by others, and only a few reported validity measures such as reliability and accuracy. 


Finally, S38 took up the question of security as a measureable dimension of software quality, examining 16 papers concerning security evaluation. 
Although S38 was the only review that took this approach, the potential of a security-informed approach to quality seems high. S38 observed a strong relationship between security evaluation studies and practical engineering concerns and found that security evaluation studies tended to use some of the same software metrics as quality studies. 
Given the large amount of work being done in software security overall, we take the lack of security studies in our quality studies corpus to be a reflection of disciplinary boundaries.  This may represent opportunities for deeper collaboration between software engineering researchers concerned with quality and computer science researchers concerned with security.

\section{Findings: Trends in Conducting and Publishing Secondary Reviews on Software Quality}
\label{sec:results}

\begin{figure*}[t]
\caption{Visualizing secondary study coverage. Each of the 75 secondary studies we identified are represented as a line. A blue dot indicates year of publication, segment color indicates perspective on quality, and segment length indicates the duration of the primary studies they examine. Y-axis labels give the study identifier and the number of studies examined by each.}
\includegraphics[width=\textwidth]{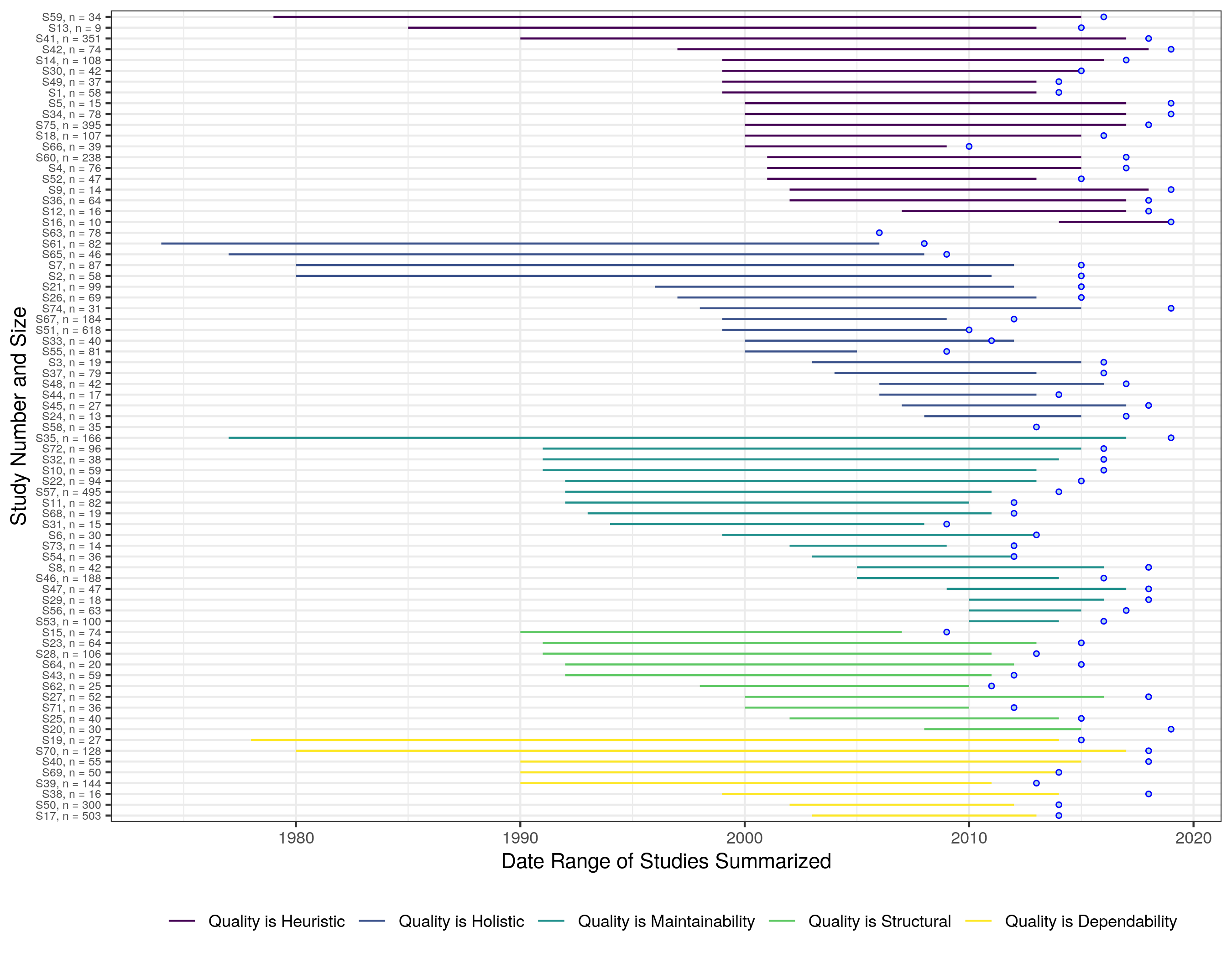}
\label{fig:pubPlot}
\end{figure*}

We analyzed the timespan coverage of the reviews we identified, the venues in which they were published, their methods, and the portion of the software development process being considered. 
Figure \ref{fig:pubPlot} offers a visualization of range of publication dates for source articles reviewed by each secondary review as well as the publication dates for each of the systematic reviews. We use the reported date range when the authors include it. When they do not report this information, we record the date range based on the date range we observe when sufficient bibliographic detail is reported; when neither is available, no date range appears. Note that this means some studies do not have a date range, and because a study might review a paper that is not officially published until later, it is possible for reviews to be published before the end of the time range covered.
The number of secondary studies has increased over the last two decades, with most studies published within the last 10 years. For each theme, at least one article in our dataset that reviewed work published in the 1990s or before. 

The secondary quality studies we identified were published most often in journals (n=49) but also appeared in conferences (n=22), symposiums (n=3), and a workshop (n=1).
Journals publishing secondary reviews were both discipline-wide and subfield-oriented. Those journals publishing more than one review are listed in Table \ref{tab:topvenues}. The broad range of venues suggests that the topic is of high interest across the field of software engineering.




The secondary analyses we identified were conducted using a range of methods. Most studies we identified described themselves as performing a systematic literature review (n=41)
and often referred to \citet{kitchenham_evidence-based_2004} and \citet{kitchenham_systematic_2009}. 
Other secondary studies identified their method as ``grounded theory,'' a ``survey,'' a ``literature survey,'' a ``meta-synthesis,'' or as a ``multivocal literature review.'' Two identified their method as simply a ``literature review'' but met our criteria because they described following a systematic search process. 

\begin{table}
\caption{Venues publishing the most software quality reviews. }
\begin{center}
\begin{tabular}{A r}
Venue & Count \\    \hline Information and Software Technology &   8 \\    Journal of Systems and Software &   8 \\    IEEE Transactions on Software Engineering &   3 \\    Empirical Software Engineering &   2 \\    Emprical Software Engineering and Measurement (ESEM) &   2 \\    IEEE Int. Conference on Software Architecture (ICSA) &   2 \\    Int. Conference on Software Technologies (ICSOFT) &   2 \\    Journal of Software Maintenance and Evolution: Research and Practice &   2 \\    
\label{tab:topvenues}
\end{tabular}
\end{center}
\end{table}

Secondary authors used a range of approaches to conduct synthesis and analysis. 
62 studies categorized their identified primary articles into thematic categories, while 58 report additional statistics beyond the thematic categories they identified. Although only three articles describes themselves as a  ``meta analysis,'' 
authors of 6 reviews were able to conduct some form of secondary analysis of the data from the studies they identified.
9 
studies offered a paragraph summary of each article they identified. 

Several practices support reproducibility in secondary reviews: dataset publication, 
by-venue online appendices, using paper-specific identifiers rather than only stating aggregate statistics such as ``3 of 32 papers,'' and inclusion of a secondary bibliography.
We found a diverse range of strategies for reporting source material.
67 studies included a list of all of their primary studies, of which 49 studies included what we describe as ``all data''---not only a bibliographic record, but also sufficient detail on the codes and traits assigned to each record that the analysis could be reproduced. Some included these materials in tabular dataset form. 
8 studies did not offer bibliographic records for the studies they identified and reported upon. 
This undermines both reproducibility and the ability of a review to act as a reading list for researchers interested in a topic.




\section{Discussion}
\label{sec:discussion}

\subsection{The Five Perspectives on Quality}

We identified five themes that reflect perspectives on quality in use by the software engineering research community. 
Taken together, these five form an ontology that can be used to classify and characterize the development of future theoretical approaches to software quality measurement. 

\textit{Quality is Heuristic} is a `clean code' perspective. Software engineering researchers with this perspective first articulate what makes some code more desirable than other code, and then attempt to embed this wisdom into problem detection algorithms. This perspective emphasizes the engineer, including their aesthetic preferences, sense of proficiency, and personal productivity. This approach treats user-oriented considerations as a desirable and assumed, but often unverified, after-effect (e.g., clean code is assumed to be faster to adapt to new requirements, assumed to have fewer bugs, and so on.). 
Similarly, the \textit{Quality is Maintainability} perspective centers the experience of developers. It also attempts to reflect the fact that modern software production is not the completion of a single bounded project. Rather, it is treated as an ongoing cycle of development and release that proceeds through time.
By contrast, the \textit{Quality is Holistic} perspective reaches across multiple perspectives and facets and seeks to be broad or all-encompassing, potentially at the expense of comparability. 
The \textit{Quality is Structural} perspective draws from classic engineering and manufacturing perspectives about defects as a predictable aspect of a product. 
Finally, \textit{Quality is Dependability}, like Quality is Maintainability, takes up a temporal perspective, but is oriented to the faults experienced when software is deployed and integrated.

\subsection{Popular Measures of Questionable Validity}

While many authors commented on the widespread use and availability of some metrics, several in our corpus also cast doubt on the validity of these very measures.  
For example, many authors remarked on the popularity of length, complexity, and cohesion and coupling metrics [S21, S28, S31, S33, S43, S63, S70, S74]. These metrics are also commonly found in metrics analysis tools. 


Although S43 found several studies that validated the relationship between complexity and faults,
several authors raised concerns about the validity and utility of commonly used measures [S21, S28, S55, S63, S74]. 
S21 found that some metrics may indicate higher or lower quality, depending on how quality is defined. 
For example, in the mapping study S55, the authors present a series of concerns about the theoretical validity of two long-standing metrics: the Lack of Cohesion of Methods 
and Coupling Between Objects 
metrics from Chidamber and Kemerer (C\&K). This is a significant critique given that 24 of the 25 empirical evaluation studies S55 identified made use of them. 
Of the five metrics S37 found to be the most common in holistic studies, we observe that concerns about theoretical validity have been raised with four of them. 
Likewise, the four metrics that S70 found were used most frequently in reliability studies had been reported to have validatity issues. 
S55 also describes theoretical validity concerns with the Lines of Code metric as a quality measure, 
given its relationship to rate-based metrics such as those examining defect density.

These issues go beyond theoretical concerns. Multiple reviews have reported inconsistent empirical results with respect to quality metrics. 
To make sense of this conflict, S21 examined multiple facets of quality (including reliability and maintainability), and found that different metrics performed better for different facets. 
Many widely used measures were gauged significant less than 50\% of the time, the least reliable being Number Of Children (NOC) and Depth of Inheritance (DIT), which in reliability studies were significant 29\% and 25\% of the time, respectively. 
Similarly, with respect to maintainability, S21 found that while some cohesion metrics offered useful results, DIT and NOC were much less reliable and were significant in only 16\% and 11\% of datasets respectively. 
These troubled metrics are all widely available through metric extraction tools [S45]. 
Although the heavy use of questionable metrics like NOC and DIT in these metric extraction tools is a liability, tool improvement represents a ready solution. 

\subsection{The Promise of Machine Learning}

Machine learning methods are increasingly being used to measure quality \cite{fernandez_del_carpio_techniques_2018} and ML studies appear in our corpus from multiple perspectives [S5, S15, S20, S23, S25, S28, S62, S71]. However, several authors found that a lack of shared reporting standards for evidence and evaluation in ML is substantially hampering progress toward the use of ML to support evidence-driven engineering [S20, S23, S71].

For example, when S71 examined the state of the evidence for ML techniques in bug prediction, they began with 208 studies but found that only 36 passed their quality criteria. The excluded articles either lacked details on the context of the study or, despite claiming prediction as their goal, only conducted correlational analysis. Of the remaining 36, only 19 reported sufficient performance data to allow direct comparison across results. 
We want to highlight the missed opportunity here. Fewer than 10\% of the empirical studies in this area articulated evaluation measures with detail sufficient to allow their results to be systematically compared to other studies. 


Future researchers considering data science techniques for assessing software quality may find the clear reports made by \citet{fernandez_del_carpio_techniques_2018} with respect to ML thresholds and model fitness to be of particular value.
Addressing these reporting issues is relatively straightforward, but requires attention from researchers as well as reviewers and editors.

\subsection{Data Sources}
Many authors remarked on the predominance of Java codebases as objects of study [S14, S21, S23, S28, S37, S52, S53, S71]. In fact, software quality research's empirical focus is even more narrow: Eclipse was the most common codebase in the studies S21, S23, and S71 reviewed, and JHotDraw was the most common codebase in the studies S52 reviewed.  
This is a clear case of convenience driving research decisions: there are abundant tools to assess Java metrics and both Eclipse and JHotDraw have freely licensed public Java codebases [S53]. The growing body of work using similar tools, metrics, and languages suggests the potential for future meta-analytical work.
However, this concentration on Java and Eclipse may be a limitation on metric generalizability since effective metrics may vary by language and the ultimate purpose of the codebase, even within the OO paradigm. 


Multiple studies remarked on the influence of the PROMISE repository, a set of public datasets released in 2005, as substantially increasing not only the amount of research investigating software quality, but also the use of ML techniques and the number of studies using public datasets [S15, S20, S21, S23, S27, S28, S37, S62, S64, S71]. 
However, as S20 describes, the use of PROMISE datasets may be problematic, given both a series of issues with their quality as well as limitations imposed by their age given that much of the data is derived from NASA projects in the 1970s-1990s. 
Several authors expressed concern that use of public data is too low 
since the results of proprietary studies are not repeatable [S21, S23, S28]. 
S28 found the number of studies using small datasets to be relatively high (33\% of their sample). 

\subsection{Toward a Theory of Software Quality}


Despite the relative maturity of software quality research as a field
software quality research is frequently disconnected and in conflict. When empirical work is failing to bear consistent fruit, it may be, as Kurt Lewin offered, that ``there is nothing as practical as a good theory.'' 

After reviewing the literature, our sense is that software quality is most typically discussed at three different levels. 
At the highest level, researchers refer to quality in terms of \textit{specific conceptual dimensions}.
These conceptual dimensions are not measurable but reflect important attempts to break down the quality mega-concept into manageable pieces. These dimensions are referred to `facets' in the ISO quality standards and will frequently include sub-concepts like reliability, dependability, and so on. Secondary reviews that treat \textit{quality as maintainability}, \textit{as holistic}, and \textit{as dependability}, tend to engage with quality at this highest level.

At the base level are \textit{direct metrics} that can be measured quantitatively. They are typically things like the number of bugs, cost of future maintenance in dollars, end-user satisfaction, and so on. Many of the reviews in our \textit{quality is structural} theme take this approach. One challenge with direct metrics is that many of the measures for which there is the most enthusiasm and agreement are very hard to measure in general or else difficult to measure until after it is too late. For example, one wants to know if software is buggy before those bugs impact production use of the software. 

As a result of this limitations, software quality researchers rely on \textit{proxy and predictive measures} which exist at a level somewhere between the two other levels. These measures typically reflect ways of writing code that engineers and practitioners believe will lead to higher quality software in terms of some other set of metrics, but which they cannot measure as easily. Our theme about \textit{quality as heuristic} captures work that focuses on these types of proxy and predictive measures.

Each of these levels offers a set of risks and challenges. Items in each level are also connected to the other levels in a tangle of untheorized and unexplored ways. We suggest that by placing empirical software quality work into this framework, we can understand how apparently contradictory research on software quality may in fact be complementary. By calling our attention to how we are conceptualizing and measuring quality, we can identify things that we have taken for granted in previous research and identify next steps.

\subsubsection{Specific Conceptual Dimensions}

There is wide recognition within software engineering research that quality is multi-faceted. It is also well known that certain types of interventions to improve quality may increase quality in one dimension while decreasing it in another (e.g., attempts to make software more reliable might decrease performance). What is foregrounded less frequently is the fact that a multi-faceted approach to conceptualizing quality means that desirably dimensions of quality can be, and often will be, in direct conflict. When it comes to software quality, we simply cannot always have it all. Future work should map out the relationships and conflicts between different facets of quality identified in this tertiary review and in the rest of the literature.

A more fundamental challenge is that individual dimensions of quality will always be operationalizable into multiple conceptually valid metrics. As we have described, there are many reasonable ways to approach measuring a dimension of quality like maintainability, reliability, or dependability. These different measures will sometimes conflict with each other. In some cases, this may reflect a lack of conceptual clarity. For example, the fact that different measures of smell and technical debt can point in such different directions is evidence that the concepts of code smell and technical debt may be overloaded. The software engineering literature has not consistently or clearly mapped specific concepts to measures. It can do so much better. Discussion of `validation' will be misleading unless we specify the underlying measure(s) against which a concept or an intervention is being validated. ML tools can allow the faster assessment of large volumes of code, but cannot relieve these challenges related to conceptual validity. 

\subsubsection{Direct Metrics}

The approach of focusing on direct measures of quality is often a feature of the \textit{quality as structural} perspective. In ways that we have already hinted at, measures of the same concept may be in conflict with each other just as concepts may be in theoretical conflict. 
This will be aggravated by the fact that measures might not nest in a clean hierarchical way due to conceptual overlap and fuzziness.

Of importance is recognition that no set of measures can be a perfect measure of any concept. No approach to bug identification or tracking will result in a complete or unbiased measure of faultiness. No survey can fully capture how end-users feel about a piece of software. Although research in the \textit{quality is structural} perspective is often engaged in a hunt for metrics that can make code ``fault-prone,'' developers will always discover ways to work to the metrics in ways that are at least partially at odds the desired underlying outcome.
When it comes to quality, there can be no true measure of ground truth. Using multiple metrics can help but is no panacea.

\subsubsection{Proxy and Predictive Measures}

Perhaps the most fraught approaches to assessing software quality rely on proxy and predictive measures. With its focus on code smells, the \textit{quality as heuristic} perspective is the most clear source of these examples. Code smells are a proxy because smelly code is not lower quality \textit{per se}. Smells are signs of a form of low quality that can be detected in advance. Software engineering research has often assumed that proxy measures like smells are good measures of quality. 

Sometimes these assumptions take the form of not clearly specifying which lower level direct metrics will be associated with a higher-level proxy. There is not consensus about what, specifically, code smells are bad for: Will smelly software be buggier? Will it lead to lower levels of end-user satisfaction? Will be it more expensive to maintain? Will it lead to unhappy engineers and higher developer turnover? 

In situations where these assumptions have been clearly articulated, they are frequently not explicitly validated in terms of direct metrics.
This is not say that proxies are misleading. For example, the presence of code smells may signal a lack of experience or formal expertise.
Practitioners who avoid smells may be those who are trained and experienced, given adequate time to complete their work thoughtfully, and have access to tools, mentorship and effective code review to support their work. 
Hence, proxy measures may have a modest relationship with varying metrics and dimensions of quality through either direct pathways or simply as markers of other advantages that may correlate to easy-to-maintain and low-fault code.

The lack of validation of widespread proxies like measures of code smells and technical debt in terms of underlying and more direct measures of code reflects the most important example of low hanging fruit in software quality research. 
Software engineering researchers can articulate the the relationships between widely used proxies and overarching concepts. They can also specify the relationships between proxies and a range of different underlying measures of quality. Each relationship can serve as a theoretical proposition and a testable hypothesis. 
In this way, scholars can begin to identify the underlying conceptual models behind software quality.
Doing so will not only help us identify what works and what does not, but also the reasons for success and failure.

\section{Threats to Validity}
\label{sec:threats}

One key limitation to this study is imposed by our method: by focusing on secondary studies, we omit lines of investigation that have not yet attracted secondary reviews. 
Further, the volume of available literature and the polysemy of relevant terms may limit our ability to be comprehensive. We have sought to mitigate this risk by using both diverse digital libraries for our preliminary search and the reverse snowballing technique. By omitting work published in predatory outlets, self published materials, and gray literature as a way to filter for quality, we may have neglected useful contributions or offered an incomplete picture of researcher interest. Given that we found a substantial volume of studies, we take this risk to be minimal, and believe it is outweighed by the gain in terms of validity. Furthermore, our finding of diverse publication venues suggests that there are many outlets for high-quality reviews.

Finally, our search and review process may have introduced some unconscious bias into our results; 
in particular the determination of which studies were concerned with contexts we gauged as too narrow for our purposes (e.g., automotive software) versus those contexts we gauged sufficiently broad (e.g., object-oriented code). 


\section{Conclusion}
\label{sec:conclusion}

This study synthesizes current knowledge about how software quality can be measured and summarizes current understanding of how the concept of quality is being defined and determined.
We found no easy answer: software quality research is a highly active field of inquiry, engaging with definitional, tactical, and strategic challenges. The field has made substantial progress but remains far from achieving consensus. 
Quality research faces multiple challenges.

Lack of validated empirical work and gaps in statistical methods and reporting is a persistent concern. ML techniques offer exciting new strategies for tackling the data intensive nature of software development, but generally require valid training data.
Studies taking on more comprehensive or holistic assessments and engaging in comparative work, if accompanied by validation, are also needed. 
Some long-standing metrics examining complexity and cohesion have demonstrated their empirical validity for specific facets of quality, but attempts to validate many others have served to illustrate the complex trade-offs engineers must make when they design, implement, and maintain software. 

Adequate reporting and weighing of evidence is key. We note that authors of secondary works were able to draw the strongest conclusions when primary authors used similar measures and consistently reported statistics that support comparison across studies, such as AUC, sample size, and correlation strength.
However, we also observed that secondary studies would frequently observe that a metric was used often without weighing the degree to which it has been validated. Given that frequent use is not evidence of correctness, reviews that do not weigh validation may contribute to a misunderstanding of the nature and relative weight of evidence.

We join S41 in observing the significance of open science practices including public, validated datasets. 
Adopting an even more more open approach may indeed be necessary for the field to make progress. Given that so many findings that are inconsistent across empirical settings, more durable conclusions may only be possible when data is sufficiently available to enable large-scale cross-project comparative work. Worryingly, \citet{mahmood_reproducibility_2018} found that of the 208 defect prediction studies identified through a widely cited SLR [S71], only 13 had drawn an attempted replication. Of those 13 studies, the replications agreed with the original in only 62\% of the cases. In that some studies drew multiple replications, 18 of the 29 replication experiments agreed with the original study. 

Software quality research addresses a vital topic of concern to our economic, political, and social life. Evidence-driven software engineering needs measurement strategies that are both actionable and valid. As development practices continue to evolve and new design challenges emerge, the challenge for the field is to remain both sufficiently grounded in the day-to-day realities of practical development and sufficiently oriented by robust theoretical work and empirical analysis to offer evidence to guide that development. 

\section*{Acknowledgments}
University of Washington librarians Jessica Albano and Mel DeSart provided helpful advice in the planning of this project. Members of the Community Data Science Collective provided excellent advice on an early draft of this project, including Jeremy Foote, Floor Fiers, Wm Salt Hale, Sohyeon Hwang, Charles Kiene, Aaron Shaw, and Nathan TeBlunthuis. Attendees at LinuxFest Northwest and the Seattle Area GNU/Linux conference provided valuable feedback throughout this project. The authors gratefully acknowledge support from the Alfred P. Sloan Foundation through the Ford/Sloan Digital Infrastructure Initiative, Sloan Award 2018-11356.

\bibliographystyle{IEEEtranN}
\bibliography{bibliography}

\renewcommand{\refname}{Secondary Literature}


\end{document}